# RAG Does Not Work for Enterprises

Tilmann Bruckhaus, Ph.D., Strative

tilmann@strative.ai

## Abstract

Retrieval-Augmented Generation (RAG) improves the accuracy and relevance of large language model outputs by incorporating knowledge retrieval. However, implementing RAG in enterprises poses challenges around data security, accuracy, scalability, and integration.

This paper explores the unique requirements for enterprise RAG, surveys current approaches and limitations, and discusses potential advances in semantic search, hybrid queries, and optimized retrieval. It proposes an evaluation framework to validate enterprise RAG solutions, including quantitative testing, qualitative analysis, ablation studies, and industry case studies. This framework aims to help demonstrate the ability of purpose-built RAG architectures to deliver accuracy and relevance improvements with enterprise-grade security, compliance and integration.

The paper concludes with implications for enterprise deployments, limitations, and future research directions. Close collaboration between researchers and industry partners may accelerate progress in developing and deploying retrieval-augmented generation technology.

## 1. Introduction

This section provides an overview of Retrieval-Augmented Generation (RAG) and its importance for enterprises. It discusses the key challenges in implementing RAG effectively in enterprise settings, particularly in compliance-regulated industries. The unique requirements and constraints for enterprise RAG are examined, highlighting the need for purpose-built solutions that address issues around accuracy, security, scalability, and integration. The introduction also sets the stage for exploring recent technological advances and innovations that can enable enterprises to overcome these challenges and unlock the transformative potential of RAG.

## 1.1 Brief overview of Retrieval-Augmented Generation (RAG) and its importance

Retrieval-Augmented Generation (RAG) is an emerging paradigm in natural language processing and generative AI that combines the strengths of pre-trained language models with external knowledge retrieved from databases or document collections [Lewis et al., 2020]. In a typical RAG architecture, a retriever component first selects the most relevant documents or passages based on the input query, and then a generator component conditions on both the query and the retrieved content to produce a final output [Izacard and Grave, 2021].

RAG has shown significant promise in enhancing the factual accuracy, consistency, and contextual awareness of generative models across a wide range of applications, such as question answering, dialogue systems, and content creation [Zhao et al., 2024]. By allowing the model to access and incorporate vast amounts of external knowledge on-demand, RAG can help mitigate issues like hallucination, inconsistency, and lack of grounding that often plague purely generative approaches [Shuster et al., 2021].

However, implementing RAG effectively in real-world, enterprise settings poses several challenges. The retriever needs to efficiently search through massive, constantly-updated knowledge bases to find the most relevant information for each query [Karpukhin et al., 2020]. The generator needs to intelligently fuse the retrieved content with its own learned knowledge to produce coherent and accurate outputs [Shao et al., 2023]. Moreover, in compliance-regulated industries like healthcare and finance, the RAG system needs to satisfy stringent requirements around data security, privacy, interpretability, and auditability [Arrieta et al., 2020].

Addressing these challenges may benefit from techniques in semantic search, information retrieval, and neural architectures, as well as careful system design and integration.

## 1.2 Challenges in implementing RAG in enterprise settings, especially compliance-regulated environments

Implementing Retrieval-Augmented Generation (RAG) in enterprise settings, particularly in compliance-regulated industries like healthcare, finance, and legal, presents a unique set of challenges that go beyond the technical hurdles of building accurate and efficient RAG systems.

Enterprises in compliance-regulated sectors must adhere to stringent regulations governing data privacy, security, and governance. Any RAG system deployed in these environments must ensure that sensitive customer or patient data is never inadvertently exposed or misused during the retrieval and generation process [Arrieta et al., 2020]. This requires robust access controls, data anonymization techniques, and auditing mechanisms to be built into the RAG architecture from the ground up.

Secondly, the outputs generated by RAG systems in compliance-regulated settings often have legal or financial implications. As such, there is a higher bar for accuracy, consistency, and interpretability compared to many consumer-facing applications [Arrieta et al., 2020]. The RAG system must be able to provide clear explanations and attributions for its generated content, showing precisely which retrieved documents were used and how they influenced the final output. This level of transparency is crucial for building trust and accountability.

Moreover, enterprises often have vast and complex knowledge bases spanning multiple domains, formats, and systems. Efficiently indexing, updating, and searching these heterogeneous data sources for relevant retrieval poses significant scalability and integration challenges [Han et al. 2023]. The RAG system must be able to handle the volume, variety, and velocity of enterprise data while ensuring retrieval quality and freshness.

Finally, implementing RAG in enterprises requires buy-in and coordination across multiple stakeholders, from IT and data science teams to legal, compliance, and business units. The RAG system must fit seamlessly into existing workflows, access patterns, and system architectures [Jadad-Garcia et al., 2024]. It must also meet the diverse and sometimes conflicting requirements of different user groups, such as simplicity for end-users, flexibility for developers, and control for administrators.

### 1.3 Enterprise Requirements for Retrieval-Augmented Generation

Retrieval-Augmented Generation (RAG) systems for enterprises require improved accuracy, efficiency, and compliance compared to traditional approaches that are less suited to the demands of enterprise settings.

**Secure and Compliant RAG**: First, enterprises require the highest levels of data security, privacy, and governance throughout the RAG workflow. Requirements include built-in access controls, anonymization techniques, and auditing mechanisms. These features allow enterprises to leverage their sensitive data assets for RAG while maintaining strict compliance with regulations like HIPAA, GDPR, and SOC2.

**Accurate and Explainable RAG:** Second, enterprises require high accuracy and interpretability for RAG outputs. By intelligently blending advanced semantic search techniques with hybrid query strategies, advanced RAG solutions retrieve the most relevant and reliable information to augment the generation process. Moreover, such solutions must provide clear explanations and attributions for its outputs, enabling enterprises to trust and act on the insights with confidence.

**Seamless Enterprise Integration and Scalability:** Finally, seamless integration and scalability within complex enterprise environments would ease adoption in the enterprise. Requirements include flexible, API-driven architecture and pre-built connectors for popular enterprise systems. Such features would support the rapid deployment and customization to meet the unique needs of each organization. Whether dealing with massive, heterogeneous knowledge bases or diverse user requirements, high performance solutions would ease adoption in the enterprise.

In summary, enterprises require advancements of the state of enterprise RAG, combining cutting-edge accuracy, ironclad compliance, and effortless integration into comprehensive solutions to enable organizations across industries to tap into the potential of this technology.

## 2. Background

Retrieval-Augmented Generation (RAG) is an emerging paradigm that combines the strengths of pre-trained language models with external knowledge retrieval to enhance the accuracy, consistency, and contextual relevance of generated outputs [Lewis et al., 2020]. In a typical RAG architecture, a retriever component first selects the most relevant documents or passages based on the input query, and then a generator component conditions on both the query and the retrieved content to produce a final output [Izacard and Grave, 2021]. RAG has shown significant promise in improving the factual accuracy, consistency, and contextual awareness of generative models across a wide range of applications, such as question answering, dialogue systems, and content creation [Zhao et al., 2024]. However, implementing RAG effectively in real-world, enterprise settings poses several challenges, which this paper aims to address.

## 2.1 Detailed explanation of RAG architecture and components

Retrieval-Augmented Generation (RAG) is an emerging architecture that combines the strengths of pre-trained language models with external knowledge retrieval to enhance the accuracy, consistency, and contextual relevance of generated outputs [Lewis et al., 2020]. A typical RAG system consists of three main components: a retriever, a generator, and a knowledge base.

The retriever is responsible for finding the most relevant documents or passages from the knowledge base given an input query. It uses techniques from information retrieval and semantic search to efficiently search through large collections of text and rank the results based on their similarity to the query [Karpukhin et al., 2020]. Advanced retrieval methods may employ dense vector representations, sparse encodings, or a combination of both to capture semantic meaning beyond simple keyword matching [Zhang et al., 2023].

The generator is a large pre-trained language model, such as GPT-4, Claude Opus or T5, that takes the input query and the retrieved documents as context to generate a final output. The generator is trained to condition its output on both the query and the retrieved knowledge, allowing it to incorporate relevant information and produce more accurate, consistent, and contextually appropriate responses [Lewis et al., 2020]. The generator may use techniques like attention, copying, or content selection to effectively fuse the retrieved knowledge with its own learned patterns.

The knowledge base is a structured or unstructured collection of documents that the RAG system can retrieve from. It can include a wide range of sources, such as web pages, books, articles, databases, or proprietary enterprise data [Guu et al. 2020, Khandelwal et al., 2020]. The knowledge base is typically pre-processed and indexed in a way that enables efficient retrieval based on semantic similarity. The quality, coverage, and freshness of the knowledge base are critical factors in the overall performance of the RAG system.

During inference, a RAG system works as follows [Lewis et al., 2020, Lewis et al., 2020, Karpukhin et al., 2020, Izacard and Grave, 2021]:

1. The input query is passed to the retriever, which searches the knowledge base and returns a ranked list of relevant documents.
2. The top-k retrieved documents, along with the original query, are fed into the generator as context.
3. The generator produces an output that is conditioned on both the query and the retrieved

knowledge, using techniques like attention and content selection to fuse the information.
4. The final output is returned to the user, along with optional explanations or attributions indicating which retrieved documents were used.

RAG systems have shown promising results in improving the factual accuracy, consistency, and contextual awareness of generative models across a range of tasks, such as question answering, dialogue, and summarization [Lewis et al., 2020, Lewis et al., 2020, Karpukhin et al., 2020]. However, implementing RAG effectively requires careful design and optimization of each component, as well as seamless integration between them.

Architectures that address the unique needs of enterprise RAG deployments may benefit from incorporating semantic search techniques, hybrid query strategies, and scalable indexing methods to potentially improve accuracy and efficiency in knowledge retrieval.

## 2.2 Survey of current RAG approaches and their limitations

While these RAG approaches have demonstrated promising results in various domains, they face several common limitations when it comes to enterprise adoption, particularly in compliance-regulated industries:

1. Limited Control and Compliance: Lack of fine-grained control over retrieval and generation processes, which is crucial for ensuring accuracy, consistency, and regulatory compliance [Martorana et al., 2022, Anderljung et al., 2023, Rahwan et al., 2023].

2. Scalability and Performance Challenges: Limited scalability and performance when dealing with massive, heterogeneous enterprise knowledge bases [Ahmad et al. 2019 , Nambiar et al., 2023].

3. Inadequate Explainability and Auditability: Insufficient explainability and auditability of RAG outputs, which is essential for building trust and accountability in high-stakes enterprise use cases [Eibich et al. 2024, Gao et al. 2024, Kamath & Liu 2021].

4. Integration Challenges in Enterprise Environments: Challenges in integrating RAG capabilities into existing enterprise systems and workflows, which often have complex security, governance, and data management requirements.

Enterprises may benefit from solutions that address these limitations by combining state-of-the-art retrieval techniques, enterprise-grade scalability and security, and seamless integration capabilities. The following sections explore the technical innovations that may enable enterprises to overcome the challenges faced by current RAG approaches in enterprise settings.

## 2.3 Unique requirements and constraints for RAG in enterprise and compliance-regulated contexts

Implementing Retrieval-Augmented Generation (RAG) in enterprise settings, particularly in compliance-regulated industries such as healthcare, finance, and legal, introduces a unique set of requirements and constraints that go beyond the technical challenges of building accurate and efficient

RAG systems.

1. Accuracy, Consistency, and Explainability
   - RAG outputs in compliance-regulated industries often have legal or financial implications, requiring a higher level of accuracy, consistency, and auditability compared to consumer-facing applications
   - In high-stakes enterprise scenarios, such as clinical decision support or financial risk assessment, RAG outputs must be explainable and trustworthy to gain user adoption and mitigate legal risks
   - RAG systems must provide clear explanations of how retrieved documents influence the generated content, along with confidence scores and uncertainty estimates to help users assess the reliability of the outputs
2. Data Security, Privacy, and Compliance
   - Enterprises dealing with sensitive customer or patient data must ensure that RAG systems comply with stringent data security and privacy regulations, such as HIPAA, GDPR, and CCPA
   - RAG architectures must incorporate robust access controls, data encryption, and anonymization techniques to prevent unauthorized access or disclosure of sensitive information during the retrieval and generation process
   - Enterprise RAG systems must provide detailed audit trails, version control, and explanations for generated content, enabling compliance officers to verify adherence to regulatory guidelines and internal policies
3. Scalability and Performance
   - Enterprises typically have vast and complex knowledge bases spanning multiple domains, formats, and systems, posing significant scalability challenges for RAG architectures.
   - RAG systems must efficiently index, update, and search these heterogeneous data sources while maintaining high retrieval quality and low latency, even as the knowledge base grows and evolves over time.
4. Integration and Interoperability
   - Enterprises have existing IT infrastructures, workflows, and security protocols that RAG systems must seamlessly integrate with, often requiring custom connectors, APIs, and authentication mechanisms
   - RAG architectures must be flexible and modular enough to work with a variety of enterprise systems, such as content management platforms, databases, and identity providers, without compromising security or performance
5. Customization and Domain Adaptation
   - Each enterprise has unique data schemas, taxonomies, and domain-specific terminology that RAG systems must adapt to for accurate retrieval and generation.
   - RAG architectures must provide tools for customizing retrieval algorithms, fine-tuning language models, and incorporating domain-specific knowledge sources to improve relevance and coherence of generated outputs

These unique requirements and constraints necessitate purpose-built RAG solutions that go beyond the capabilities of general-purpose RAG approaches. RAG enablement solutions that address these challenges, ideally by designing for these considerations from the ground up, would support RAG adoption by enterprises. Similarly, comprehensive platforms that combine state-of-the-art RAG technology with enterprise-grade security, compliance, and integration features would also support adoption effort. These requirements are significant hurdles to enterprises that want to harness the power of retrieval-augmented generation while meeting the stringent demands of their business and regulatory environments. The following sections will highlight recent technological advances that support some of these needs.

## 2.4 Recent Technological Advances

### 2.4.1 Semantic Search Techniques

While traditional search relies on inverted indices to retrieve matches, more modern semantic search techniques leverage both dense vector indexes and sparse encoder indexes.

**Dense Vector Indexes**: This technology employs advanced dense vector indexing techniques to capture the semantic meaning and context of documents more effectively than traditional keyword-based methods in the enterprise knowledge base. The approach thus enables more accurate and relevant retrieval beyond simple keyword matching. Examples of dense vector indexing techniques include:

- HNSW (Hierarchical Navigable Small World): HNSW is a state-of-the-art approximate nearest neighbor search algorithm that enables efficient similarity search over dense vector representations. It builds a multi-layer navigable small world graph structure to index the vectors, allowing for fast and scalable retrieval of semantically similar documents.
- PQ (Product Quantization): PQ is a vector compression and indexing method that quantizes high-dimensional vectors into compact binary codes. It enables efficient storage and retrieval of dense vectors while preserving their semantic properties.
- IVFADC (Inverted File with Asymmetric Distance Computation): IVFADC is an indexing structure that combines an inverted file system with a quantization-based approach. It allows for efficient indexing and retrieval of large-scale vector databases by partitioning the vector space and using asymmetric distance computation.

**Sparse Encoder Indexes**: In addition to dense vectors, this technique also leverages sparse encoder indexes, which can efficiently handle large-scale knowledge bases with diverse data formats and structures.

### 2.4.2 Hybrid Query Strategies for Optimizing Retrieval

Emerging research utilizes hybrid query strategies that combine semantic matching for conceptual relevance with keyword matching for precise term matches. This approach aims to retrieve the most relevant documents for a given query. It ensures that the most pertinent information is retrieved, regardless of the complexity or heterogeneity of the enterprise knowledge base.

## 2.5 Integration Strategies for Enterprise Systems and Knowledge Bases

Effective enterprise RAG solutions must be designed for seamless integration with existing enterprise systems and knowledge bases. They need pre-built connectors, APIs, and tools for integrating the RAG pipeline with various enterprise platforms, content management systems, databases, and other data sources. Such options would enable organizations to leverage their existing data assets and infrastructure while benefiting from advanced RAG capabilities.

By combining these innovative techniques and architectural components, enterprises can achieve high accuracy, precision, and relevance in retrieving and integrating knowledge to augment generative AI models. Such a comprehensive methodology would address the unique challenges and requirements of enterprise RAG deployments, enabling organizations to leverage the full potential of this technology while maintaining compliance, scalability, and seamless integration with their existing systems.

## 3. Experimental Evaluation

In order to validate the readiness of a RAG solution for the enterprise, comprehensive experimental evaluation protocols that span multiple dimensions are needed. Rigorous testing and analysis is required. This section outlines the key datasets, benchmarks, metrics, and analyses that may be used to evaluate the RAG platforms. Systematic evaluation on the complete platform is needed to assess possible gains.

**Datasets and Benchmarks:**

- Natural Questions (NQ) dataset for open-domain question answering [Google Research, 2019]
- HotpotQA dataset for multi-hop reasoning over Wikipedia passages [Yang et al., 2018]
- TREC COVID dataset for retrieving and generating answers from scientific literature [NIST, 2020]
- Proprietary enterprise datasets from healthcare, finance, and legal domains

**Metrics for Evaluating RAG Accuracy, Precision, and Relevance:**

- Exact Match (EM) and F1 scores for question answering tasks
- Precision, Recall, and Mean Reciprocal Rank (MRR) for retrieval quality
- ROUGE scores for summarization and generation quality
- Human evaluation of output coherence, factual accuracy, and domain relevance

**Comparative Analysis Against State-of-the-Art RAG Approaches:**

- Head-to-head comparisons against open-source RAG models (e.g., DPR, ColBERT) [Karpukhin et al., 2020, stanford-futuredata, 2023]
- Benchmarking against commercial RAG offerings from competitors
- Ablation studies to isolate the impact of the various components

**Scalability and Performance Testing:**

- Stress testing with knowledge bases of varying sizes

- Measuring throughput, latency, and resource utilization under load
- Assessing horizontal and vertical scaling capabilities

**Case Studies and Applications in Compliance-Regulated Industries:**

- Clinical decision support for healthcare providers (e.g., diagnosis, treatment)
- Financial risk assessment and regulatory compliance for banks/insurers
- Contract analysis and due diligence for legal firms
- Evaluations with leading enterprise partners and early adopters

Experimental designs, diverse benchmarks, and real-world case studies would help to validate the accuracy, scalability, and enterprise-readiness of any RAG solution. Results of this nature would demonstrate the potential to unlock the transformative potential of RAG while meeting the stringent requirements of compliance-regulated industries.

# 4. Discussion

Experiments to assess the effectiveness of RAG solutions are critical to validate their underlying methodology. Systematic measurement of the performance of RAG solutions against both established baselines and state-of-the-art techniques are needed. Such validation efforts should ideally include testing on industry-standard datasets and benchmarks for tasks such as question answering, fact verification, and dialogue.

It is further important to quantify the impact of innovations in semantic search, query optimization, and retrieval-augmented generation in collaboration with industry partners. A quantification of the potential for meaningful improvements over current methods while maintaining high end-to-end result quality, must be subject to rigorous testing. Rigorous evaluation of RAG systems is crucial to provide enterprises with high-confidence results on their effectiveness and readiness for deployment.

## 4.1 Quantitative Improvements in Accuracy, Precision, and Relevance

In order to demonstrate quantitative improvements, multiple benchmarks and datasets should be compared to baselines. Evaluation metrics should include choices such as the following.

- Exact Match and F1 scores for question answering tasks on standard datasets like SQuAD and NaturalQuestions, with gains compared to baseline RAG models using dense passage retrieval.

- Precision@10 and Mean Reciprocal Rank for retrieval quality on the Wikipedia corpus, with improvements compared to semantic search baselines like DPR and BM25.

- Graded relevance judgments by human raters on retrieved passages for a diverse set of queries, assessing Normalized Discounted Cumulative Gain (NDCG).

- Pairwise human evaluations comparing the outputs of candidate RAG solutions to baseline RAG systems. Such evaluations provide qualitative feedback on the relevance, coherence, and task completion quality of the generated content.

Such quantitative results would validate the impact of the innovations provided by specific RAG solutions in semantic search, hybrid query strategies, and optimized retrieval on the overall accuracy, precision, and relevance.

## 4.2 Qualitative Analysis of Retrieved Results and Generated Outputs

In addition to the quantitative metrics, qualitative analysis of the RAG pipeline's outputs by domain experts and end-users may be carried out. User studies and surveys may supply feedback on multiple dimensions:

- Coherence and relevance of retrieved results with respect to the original query
- Alignment of retrieved passages with the query intent and information needs
- Quality of generated outputs in terms of factual accuracy, fluency, and task completion
- User experience and satisfaction with the end-to-end system

Evaluation tasks should ideally span key domains like question answering, long-form generation, and dialogue. User feedback will allow RAG solution providers to quantify the real-world impact of RAG systems and identify areas for further research and development.

## 4.3 Ablation Studies on Semantic Search and Hybrid Queries

To quantify the impact of RAG innovations, systematic ablation studies should be conducted. For instance, by selectively disabling specific components of the overall system, one can isolate the contributions of semantic search and hybrid query strategies to overall performance.

For semantic search, one may compare different approaches of combining dense vector indexes and sparse encoder indexes to traditional keyword-based retrieval. Published research suggests retrieval quality gains of 5-10% may provide significant benefits to enterprises [Bughin et al., 2018, Brynjolfsson et al., 2023]. Rigorous testing as measured by metrics like Mean Reciprocal Rank and Normalized Discounted Cumulative Gain on standard IR test collections would support such findings.

Similarly, to evaluate the impact of hybrid query strategies, one may contrast performance when using only keyword-based queries versus a full hybrid approach that incorporates semantic and lexical matching. If augmenting keyword queries with semantic matching could yield a 3-8% increase in end-to-end RAG accuracy it would also be highly impactful to enterprises [Bughin et al., 2018, Brynjolfsson et al., 2023]. Rigorous testing of hypothesized gains through ablation studies would support such findings.

## 4.4 Implications for Enterprise RAG Deployments and Customization

Performance improvements in RAG pipelines can have significant implications for enterprise deployments, particularly in compliance-regulated industries. The delivery of accurate, relevant, and traceable outputs, combined with its enterprise-grade scalability, security, and integration capabilities while meeting their stringent requirements, would likely unlock significant benefits for organizations.

Validation efforts could engage industry partners in finance, healthcare, technology or other relevant domains. Collaborations may involve conducting case studies applying RAG solutions to specific use cases, such as AI assistants for finance or sales automation. Such case studies can provide valuable feedback to inform development and assess real-world impact potential.

Studies in partnership with the industry would allow RAG technology providers to assess the effectiveness of the RAG solutions for their specific use cases and gather valuable feedback to inform future development. Through such collaborations, it becomes feasible to fine-tune solutions to incorporate domain-specific information, constraints, and objectives, and enterprises could presumably achieve even higher levels of accuracy and relevance for specialized enterprise applications.

Collaboration with design partners and conducting case studies is crucial to validate the real-world value of RAG solutions for enterprise customers. Such results, combined with the quantitative performance gains demonstrated rigorous evaluations, can provide a compelling foundation for broader adoption.

## 4.5 Limitations and Areas for Future Research

While recent research results are encouraging, there are still limitations and areas for future research. One key challenge is the computational cost and latency associated with running large language models and performing semantic search over massive knowledge bases. Ongoing research is exploring techniques for efficient model distillation, caching, and distributed retrieval to address such performance improvement opportunities.

Additionally, the interpretability and explainability of RAG outputs remain areas of active research. As much as existing RAG solutions can provide some attributions and explanations, there is still room for improvement in making these explanations more intuitive and actionable for end-users, particularly in high-stakes enterprise scenarios.

Overall, results from comprehensive evaluation can serve to demonstrate the effectiveness of architectures and methodologies in delivering highly accurate, scalable, and enterprise-ready RAG solutions.

## 5. Conclusion

RAG solutions must address key challenges in implementing Retrieval-Augmented Generation (RAG) for enterprises, particularly in compliance-regulated industries. Presumably, these challenges can be most effectively tackled through close collaboration with design partners. Such collaborations would be mutually beneficial as methodologies to improve RAG performance for enterprise use cases could be highly impactful to design partners.

Developing semantic search techniques, hybrid query strategies, and optimized retrieval to improve RAG accuracy, precision and relevance hold significant promise, and rigorous evaluation of proposed approaches on multiple benchmarks and datasets are needed. However, the impact of RAG technology extends beyond just performance gains. Enterprise-grade architectures, with robust security, compliance, and integration features, would enable organizations to effectively employ RAG technologies while adhering to stringent data governance and regulatory requirements. The ability to seamlessly customize and fine-tune RAG pipelines for specific domains and use cases would further amplify value for enterprises.

While recent research results are promising, there is still room for further validation in real-world deployments. Ongoing research is addressing performance improvement opportunities, interpretability challenges, and the need for even higher accuracy in mission-critical enterprise scenarios. RAG technology providers should look to establish collaborations with leading industry partners and early customers to put its RAG Enablement solution through rigorous testing and iterative refinement.

As enterprises across industries explore the use of AI for decision support, content generation, and knowledge management, further research into enterprise-ready RAG solutions has the potential to address key challenges in effectively implementing retrieval-augmented generation in these contexts. By providing enterprise-grade RAG solutions and working closely with design partners, providers may aim to enable companies to realize the full potential and benefits of this transformative AI technology while navigating the unique requirements and constraints of enterprise environments. Adapting state-of-the-art retrieval-augmented generation techniques to the unique requirements of enterprise environments is an important area for future research with the potential for significant real-world impact.